# Phase Diagram of Collective Motion of Bacterial Cells in a Shallow Circular Pool

Jun-ichi Wakita, Shota Tsukamoto, Ken Yamamoto, Makoto Katori, and Yasuyuki Yamada

*Department of Physics, Chuo University, Bunkyo, Tokyo 112-8551, Japan*

The collective motion of bacterial cells in a shallow circular pool is systematically studied using the bacterial species *Bacillus subtilis*. The ratio of cell length to pool diameter (i.e., the reduced cell length) ranges from 0.06 to 0.43 in our experiments. Bacterial cells in a circular pool show various types of collective motion depending on the cell density in the pool and the reduced cell length. The motion is classified into six types, which we call random motion, turbulent motion, one-way rotational motion, two-way rotational motion, random oscillatory motion, and ordered oscillatory motion. Two critical values of reduced cell lengths are evaluated, at which drastic changes in collective motion are induced. A phase diagram is proposed in which the six phases are arranged.

## 1. Introduction

Various types of spatiotemporal behavior of collective organisms have been observed in the flocking of birds, the swimming of schools of fish, the marching of social insects, the migration of bacteria, the walking of pedestrians, and so forth.[1-9] They have attracted the attention of many researchers in not only biological sciences but also physical and mathematical sciences. In particular, qualitative differences between individual behavior and the collective motion of organisms remind many statistical physicists of phase transitions and critical phenomena in strongly correlated systems of atoms and spins. We expect that the collective behavior can be understood independently of any biological and ecological details of individual organisms. With the recent development of computer technology, researchers have attempted to simulate the collective behavior of organisms using model systems consisting of a large number of self-propelled units.[10-14]

In our series of work on bacterial colonies growing on the surface of semisolid agar plates, both microscopic and macroscopic aspects of pattern formation have been studied. On the basis of microscopic observations of each growing colony, we showed that cell motility and multiplication that drive colony growth depend on the difference in bacterial species, as reported for *Bacillus subtilis*,[15] *Proteus mirabilis*,[16,17] and *Serratia marcescens*.[18] By macroscopic observations, however, we have clarified that the morphology of growing colonies depends only on environmental conditions controlled by the agar concentration $C_a$ and the nutrient concentration $C_n$.[15-18] Here, varying $C_a$ affects cell motility, and varying $C_n$ to some extent affects bacterial growth and proliferation rate. Many patterns reported in our study of bacterial colonies are also found in pattern-forming phenomena in physical and chemical systems. For instance, diffusion-limited aggregation (DLA)-like pattern,[19,20] Eden-like pattern,[21,22] concentric-ring pattern, and dense branching morphology (DBM) pattern[23] are observed not only in bacterial colonies but also in crystal growth and viscous fingering. The universality of these patterns in organisms and inorganic substances suggests the existence of a common mechanism that is independent of the microscopic details of the systems.

In this study, we focus on the collective motion of bacterial cells in a shallow circular pool put on the surface of a semisolid agar plate. The diameter of the pool is arranged to be not so large compared with the length of each bacterial cell, so that cell motility along the brim of the pool is strongly restricted by the pool size. The observation time is shorter than the cell division cycle, and thus the number of bacterial cells in the pool does not increase during each observation period. Therefore, the stationary motion of bacterial cells is observed in each pool. In our experiments, we prepared shallow circular pools of approximately the same diameter, but made the average length $l$ of individual cells and the cell density $\rho$ in a pool be different from each other. We expect that the difference in $l$ will change the cell motility and the difference in $\rho$ will change the interaction between bacterial cells moving in each pool. The aim of this study is to classify the collective motion of bacterial cells by systematically changing $l$ and $\rho$.

Throughout this experiment, we used the *B. subtilis* wild-type strain OG-01. Bacterial cells of this strain are rod-shaped with peritrichous flagella and swim straightforward in water by bundling and rotating their flagella. *B. subtilis* colonies grow on the surface of semisolid agar plates by cell motility and multiplication, and typically exhibit five different patterns depending on the two environmental parameters $C_a$ and $C_n$.[15] In particular, under the condition that $C_a$ is intermediate (7 g/L < $C_a$ < 8.5 g/L) and $C_n$ is high ($C_n$ > 10 g/L), the growing front of the colony repeatedly advances (in the migration phase) and rests (in the consolidation phase). As a result, the colonies form concentric-ring patterns.[15] The bacterial cells at the growing front of a concentric-ring pattern have been observed to repeat elongation and contraction, synchronizing with the periodic colony growth.

This periodic change in cell length at the growing front enables us to control the length of bacterial cells used in our experiments, as explained in the following. When we prepare a pool at the growing front in the migration phase, the average length of the bacterial cells trapped in the pool is small. On the other hand, when we set it at the growing front in the consolidation phase, we have a pool in which longer bacterial cells are swimming. In addition, the number of bacterial cells trapped in a pool becomes varied because of the spontaneous local fluctuation in cell density at the growing front. Therefore, we need to prepare a sufficiently large number of shallow pools in the vicinity of the growing front of a concentric-ring pattern of a bacterial colony. We used glass beads to make shallow pools on the agar surface.

This paper is organized as follows. In Sect. 2, we show our experimental setup. Section 3 is devoted to our experimental results. Discussion and future problems are given in Sect. 4.

## 2. Experimental Procedure

A solution containing 5 g of sodium chloride (NaCl), 5 g of dipotassium hydrogen phosphate ($K_2HPO_4$), and 30 g of Bacto-Peptone (Becton, Dickinson and Company, Franklin Lakes, NJ, USA) in 1 L of distilled water was prepared. The environmental parameter $C_n$ was set to 30 g/L by adjusting the concentration of Bacto-Peptone. Then, the solution was adjusted to pH 7.1 by adding 6 N hydrochloric acid (HCl). Moreover, the solution was mixed with 8.3 g of Bacto-Agar (Becton, Dickinson and Company), which determines the softness of a semisolid agar plate. The environmental parameter $C_a$ was set to 8.3 g/L by adjusting the concentration of Bacto-Agar. The environmental condition realized by these $C_a$ and $C_n$ values gives a typical concentric-ring pattern of *B. subtilis* colonies.[15] The mixture was autoclaved at 121 °C for 15 min, and 20 ml of the solution was poured into each sterilized plastic petri dish of 88 mm inner diameter. The thickness of the semisolid agar plates was about 3.2 mm. After solidification at room temperature for 60 min, the semisolid agar plates were dried at 50 °C for 90 min.

3 μl of the bacterial suspension with an optical density of 0.5 at a wavelength of 600 nm was inoculated on the surface of each semisolid agar plate. The optical density of 0.5 corresponds to a bacterial density of about $10^4$ cells per μl.[15] The semisolid agar plates were left at room temperature for about 60 min to dry the bacterial suspension droplet. Thereafter, they were incubated in a humidified box at 35 °C and 90% RH.

The lag-phase period was about 7 h during which bacterial cells at the inoculation spot grew and multiplied by cell division but they did not migrate. Then, the first migration started and two-dimensional colony expansion was observed. About 2 h later, they stopped migrating and entered the first consolidation phase. They did not move but underwent cell division actively for about 5 h. Afterwards, they showed the migration phase and consolidation phase alternately. When the third migration phase or third consolidation phase were started in colony growth, we scattered glass beads of 50 ± 2 μm diameter (SPM-50, Unitika, Osaka) in the vicinity of the growing front of a concentric-ring pattern shown in Fig. 1(a). Figure 1(b) shows the situation in which the agar surface is deformed by a glass bead. Then, the beads were removed from the agar surface using adhesive tape. Finally, circular pools were produced on the surface, as illustrated in Fig. 1(c), in which bacterial cells under a glass bead were trapped and swimming. The pool depth was about 1 μm, which was much smaller than the pool diameter $d$ = 39 ± 6 μm. The thickness of each bacterial cell was about 0.5 μm, which was approximately the same scale of the pool depth. As a result, two-dimensional motions of bacterial cells were realized in the pools. No bacterial cells coming in or getting out from the pool were observed during the observation periods, so that the number of bacterial cells was unchanged. Since the water was always supplied to a pool from the agar surface, we were able to observe cell motions for relatively long time periods, which were typically more than 10 min.

The motions of bacterial cells were video-recorded using a high-speed microscope (VW-9000, Keyence, Osaka) linked to an optical microscope (DIAPHOT-TMD, Nikon, Tokyo). Cell length and pool diameter were measured manually from snapshots of the video using photo-editing software (Photoshop, Adobe, San Jose, CA, USA). The cell density in each pool was calculated by dividing the total area of bacterial cells in a circular pool by the area of the pool. Here, the areas were evaluated from the binary image of a snapshot using image analysis and measurement software (Cosmos32, Library, Tokyo).

## 3. Experimental Results
### 3.1 Six types of collective motion

We defined $l$ as the average cell length in a circular pool and the cell density $\rho$ as the average ratio of the cell area in the pool to the total pool area. Depending on $l$ and $\rho$, drastic changes in the collective motion of bacterial cells were observed. We have classified the collective motion into six types on the basis of their dynamical characteristics. Each type of motion was clearly different from the others and the types were apparent when looking at the movie. The six types of motion are explained below.

*Random motion.* In the pool in which the cell length is small ($l < 5$ μm) and the cell density is low ($\rho < 0.3$), bacterial cells showed the tendency to swim along the brim of the pool. They freely changed their directions of motion along the brim and crossed the center of the pool very frequently. Thus, they moved around two-dimensionally inside the pool. Their motions seemed to be disordered. Therefore, we defined it as random motion. Figure 2(a) shows a snapshot of their random swimming.

*Turbulent motion.* When $\rho$ became higher than that for the random motion, short bacterial cells filled the pool and moved around two-dimensionally inside the pool. In this situation, their motions along the brim of the pool were no longer identifiable and seemed to be turbulent flow [see Fig. 2(b) for a snapshot]. The creation and annihilation of vortices were observed. We defined it as turbulent motion.

*One-way rotational motion.* On the other hand, when $l$ became slightly larger than that for the random motion, the motions of bacterial cells drastically changed. Few bacterial cells were found to cross the center of the pool and the motions became one-dimensional along the brim of the pool. Most of them showed a one-way rotational motion along the brim. In this paper, the rotational direction has been defined as viewed from above the pool. We found that the rotational directions were counterclockwise along the brim, as indicated by the arrow in Fig. 2(c). Even if a few bacterial cells changed their directions to clockwise, they immediately turned to counterclockwise motion. The origin of this strong chirality is unclear, but we consider that it is caused by some biological characteristics such as the asymmetric rod shape of bacterial cells and the fixed rotational direction of flagella. We defined such behavior as one-way rotational motion.

*Two-way rotational motion.* As $\rho$ increased gradually from that for the one-way rotational motion, a circular pool became more crowded with bacterial cells from the outer region to the inner region. At a certain $\rho$ (about 0.3), a two-way rotational motion appeared, as indicated by arrows in the snapshot shown in Fig. 2(d). The bacterial cells in the outer region swam counterclockwise along the brim of the pool, while the bacterial cells in the inner region swam clockwise. Thus, they kept their motions one-dimensional along the brim. Even if a few bacterial cells in each region changed their moving directions to the opposite ones, they immediately changed their directions back to the original ones. The boundary between the two regions could be recognized clearly in a movie, although the bacterial cells sometimes moved to a different region. We defined such behavior as two-way rotational motion.

*Random oscillatory motion.* When $l$ became larger than the values mentioned above, the bacterial motions drastically changed again. The bacterial cells became unable to swim along the brim of a circular pool and showed back-and-forth movement inside the pool. In particular, when $\rho$ was low, such an axial oscillatory motion of bacterial cells seemed to be individually random (see Fig. 2(e)). Therefore, we defined it as random oscillatory motion.

*Ordered oscillatory motion.* On the other hand, when $\rho$ was high, the axial oscillatory motion of bacterial cells exhibited an interesting quasi-periodic behavior over time. For a long time duration, the axial directions of rod-shaped bacterial cells differed from each other, as shown in the right picture of Fig. 2(f) (in the disordered state). Once in a period, however, they were arrayed as shown in the left picture of Fig. 2(f) (in the ordered state). The time period was not definite and sometimes they rotated counterclockwise while maintaining the ordered state. We defined the behavior as ordered oscillatory motion.

*3.2 $(l, \rho)$ and $(\lambda, \rho)$ plots*

We analyzed the collective motion of bacterial cells in 117 circular pools. Figure 3 shows plots of the six types of observed motion. Here, the horizontal axis is $l$ and the vertical axis is $\rho$. The cell length $l$ will affect cell motility along the brim of a pool. That is, as $l$ increases, the restriction on cell motility becomes stronger. On the other hand, $\rho$ will affect the interactions between bacterial cells. As $\rho$ increases, the rod-shaped cells tend to align in the same direction in their monolayer motions.

The groups discriminated by the six types of symbol partly overlapped inside the dashed circles in Fig. 3. This ambiguity is due to the pool size distribution. Figure 4 shows the distribution of pool diameters $d$, which are measured for all the observed pools. It was found that $d$ widely ranges from 25 to 52.5 μm. Since the restriction to bacterial motions by the pool size is represented by the relative cell length compared with the pool size, we introduced the reduced cell length defined by $\lambda = l/d$ and replotted on the $(\lambda, \rho)$ plane as shown in Fig. 5. The overlapping areas shown by the dashed circles in Fig. 3 disappeared in the $(\lambda, \rho)$ plots shown in Fig. 5. Furthermore, the existence of two critical values of $\lambda$ was clarified in the $(\lambda, \rho)$ plots. Random motion and turbulent motion were observed only in the region $\lambda < \lambda_{C1} = 0.1$, and random oscillatory motion and ordered oscillatory motion were found only in the region $\lambda > \lambda_{C2} = 0.2$. One-way rotational motion and two-way rotational motion were realized only in the intermediate region $\lambda_{C1} < \lambda < \lambda_{C2}$.

*3.3 Turbulent motion in a large circular pool*

We studied the collective motion of bacterial cells in a large circular pool, which was made by using large glass beads of 100 ± 5 μm diameter (SPM-100, Unitika, Osaka). When $d =$ 89 μm, turbulent motion was observed. Here, $l = 5.9$ μm and $\rho = 0.9$, and $\lambda = 5.9/89 = 0.07$. As shown in Fig. 3, the observed turbulent motion was located at the area between two-way rotational motion and ordered oscillatory motion in the $(l, \rho)$ plots. By using the $(\lambda, \rho)$ plots, this turbulent motion was located in the proper region $\lambda < \lambda_{C1}$, as shown in Fig. 5.

*3.4 Phase diagram of collective motion*

On the basis of the $(\lambda, \rho)$ plots shown in Fig. 5, we propose a phase diagram (Fig. 6) of the collective motion of bacterial cells in a shallow circular pool. The vertical phase boundaries indicated by the shaded lines correspond to the critical values $\lambda_{C1} = 0.1$ and $\lambda_{C2} = 0.2$, at which the collective motion of bacterial cells are drastically changed. The horizontal phase boundaries indicated by the shaded areas remain uncertain owing to the insufficient amount of data. This diagram is useful for discussing the classification of the dynamical characteristics of the collective motion of bacterial cells.

## 4. Discussion and Future Problems

Here, we discuss the differences in the dynamical characteristics of the six phases using our phase diagram (Fig. 6). When $\lambda$ is less than $\lambda_{C1} = 0.1$, the motion of each bacterial cell in a circular pool seems to be free from any geometric restriction by the pool size. In particular, when $\rho$ is low, the random motion is observed. When $\rho$ becomes higher, the motions of bacterial cells seem to make turbulent flow. The change in motions resembles the transition from gas to liquid, and we expect that some hydrodynamical descriptions[24-26] will be possible in the region $\lambda < \lambda_{C1}$.

When $\lambda$ becomes larger than $\lambda_{C1} = 0.1$, bacterial cells become able to swim in the same direction along the brim of a pool. When $\rho$ is low, one-way rotational motion is observed. As we mentioned in Sect. 3.1, we consider that the counterclockwise motions are attributable to some biological characteristics of the present species of bacterial cells. For example, the bacterial species *Paenibacillus vortex* forms a rotating droplet at the branch tips of a growing colony, in which many bacterial cells are moving together in a correlated manner. Within a single colony, both clockwise and counterclockwise motions are observed.[27] For chiral and rotating colonies, see Sect. 4.1.5 in the monograph by Vicsek.[1] In our study, there is a transition from one-way rotational motion to two-way rotational motion at $\rho \simeq 0.30$, as shown in the $(\lambda, \rho)$ phase diagram. If we define $d'$ as the diameter of the circular boundary between the outer and inner regions, we find $l/d' \simeq 0.2 = \lambda_{C2}$. To see the hydrodynamical effect in a pool, we added colloid particles of 1.00 μm diameter (Polybead, Polysciences, Warrington, PA, USA) in the pool with bacterial cells. When $\rho < 0.30$, the colloid particles were driven in the clockwise direction in the outer region as a reaction to the counterclockwise motion of bacterial cells, but they were not driven in the inner region of the pool where no bacterial cells exist. This suggests that the hydrodynamical interaction between bacterial cells works at short distances. When $\rho > 0.30$, the colloid particles added in the pool were driven in the clockwise direction in both the inner region where bacterial cells move clockwise and the outer region where they move counterclockwise. This implies that, in two-way rotational motion, the hydrodynamical interaction caused by the motions of bacterial cells is relatively long-ranged. The present system in the region $\lambda_{C1} < \lambda < \lambda_{C2}$ can be regarded as a microscopic realization of the traffic currents of self-propelled particles on a circle.[28-30] The statistical-mechanical quantities such as the velocity correlation functions studied in traffic models[30] should be measured for the rotational motion of bacterial cells. The mechanism of the spontaneous formation of two-lane flow as the particle density increases is an interesting future problem.

When $\lambda > \lambda_{C2} = 0.2$, the cell movement along the brim of a pool is much restricted by the pool size. As a result, each bacterial cell shows an axial oscillatory motion inside the pool. In particular, random oscillatory motion is observed when $\rho < 0.80$, while ordered oscillatory motion is observed when $\rho > 0.80$. We expect that the statistical mechanics models of interacting oscillators studied by Kuramoto[31] will be useful for explaining such oscillatory motions observed in the present biological systems.

As mentioned above, it was found that bacterial cells show various types of collective motion depending on $\lambda$ and $\rho$, although they are confined in a microscopic-scale pool where the environmental condition $C_n$ is considered to be uniform. On the other hand, in our previous studies of bacterial colony formation, macroscopic-scale colonies showed various morphologies depending on the initial $C_a$ and $C_n$. However, the nutrient concentration field is not uniform around growing colonies in which the cell motility and multiplication locally depend on the field. It is expected that the collective behavior of bacterial cells in microscopic-scale pools is hierarchically related to that in macroscopic-scale colonies. In particular, the colony growth of DBM pattern by *B. subtilis* is driven by the collective motion of active bacterial cells at the tips of growing branches. The behavior of active bacterial cells is similar to that in a shallow circular pool. Clarifying the hierarchical structures behind colony formation is our future problem.

At present, the dynamics of the collective motion of bacterial cells in a shallow circular pool has not yet been elucidated. More extensive experiments are needed to elucidate the dynamics of each phase of collective motion classified in our phase diagram (Fig. 6). We will report the results of further studies in the near future. We hope that the above-mentioned open problems will also be studied theoretically by hydrodynamical[8,24-26] and stochastic[28-31] modeling.

**Acknowledgment**

We would like to thank M. Matsushita, Y. Yamazaki, H. R. Brand, and H. Chaté for valuable discussions. JW is supported by a Chuo University Grant for Special Research and by a Grant-in-Aid for Exploratory Research (No. 15K13537) from the Japan Society for the Promotion of Science (JSPS). KY is supported by a Grant-in-Aid for Young Scientists (B) (No. 25870743) from JSPS and MK is supported by a Grant-in-Aid for Scientific Research (C) (No. 26400405) from JSPS.

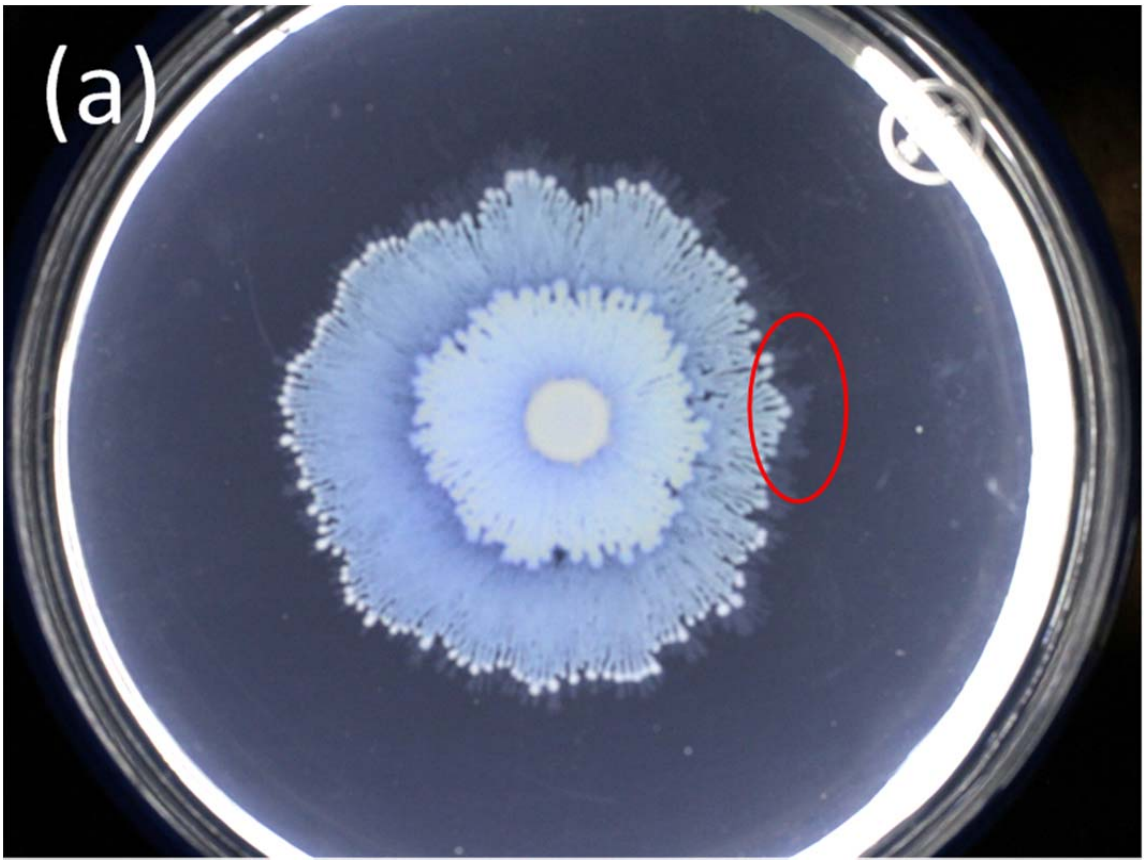

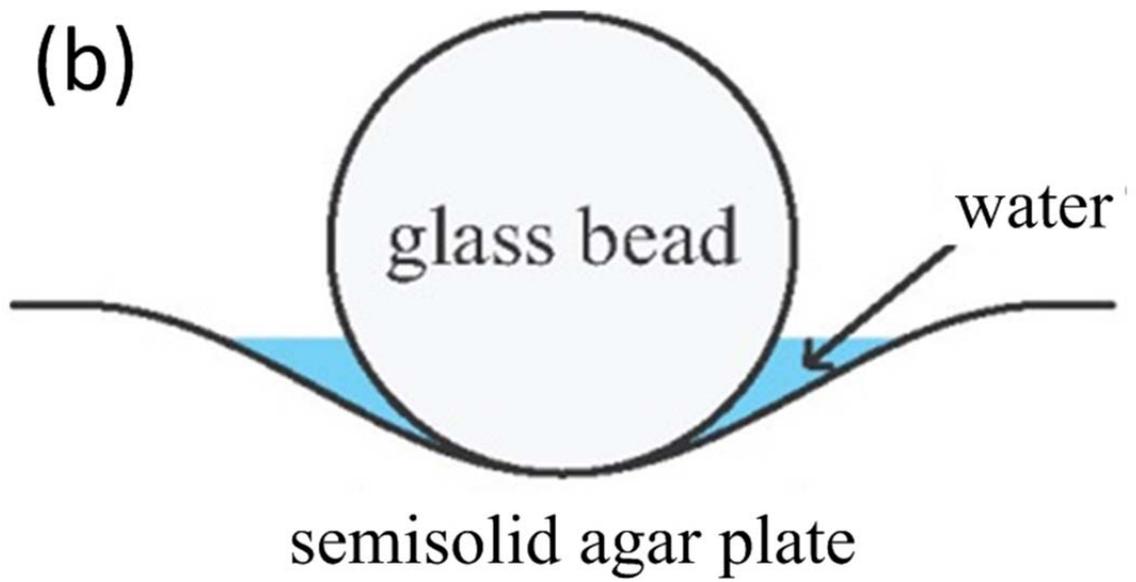

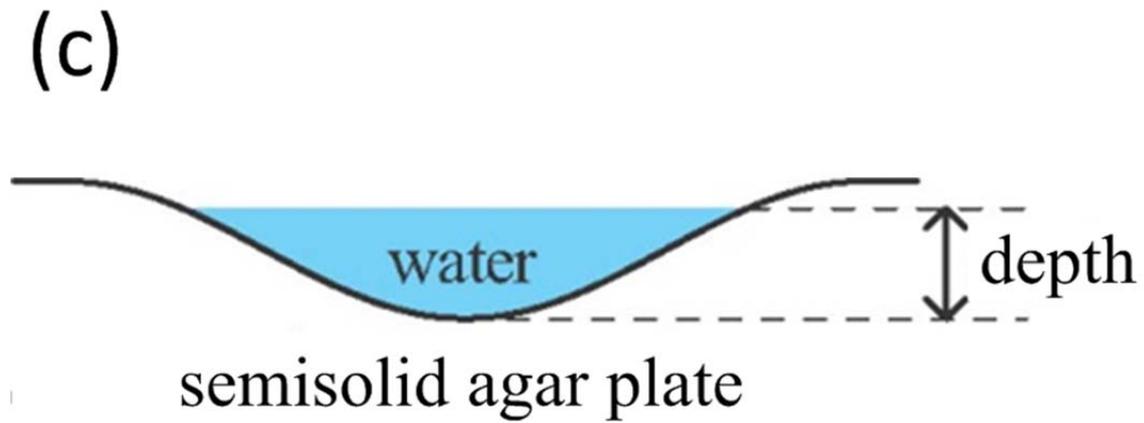

Fig. 1. (Color online) (a) A concentric-ring pattern of *B. subtilis* colonies when the growing front just starts the third migration phase. The circled area shows the region where glass beads of 50 μm diameter are scattered. (b) Illustration of a circular pool before removing a glass bead. (c) Illustration of a circular pool after removing a glass bead. The depth of the pool is about 1 μm. Bacterial cells are trapped in this circular pool.

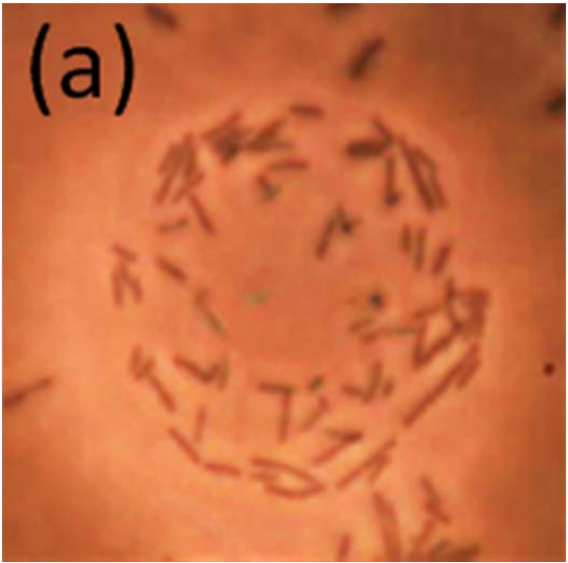

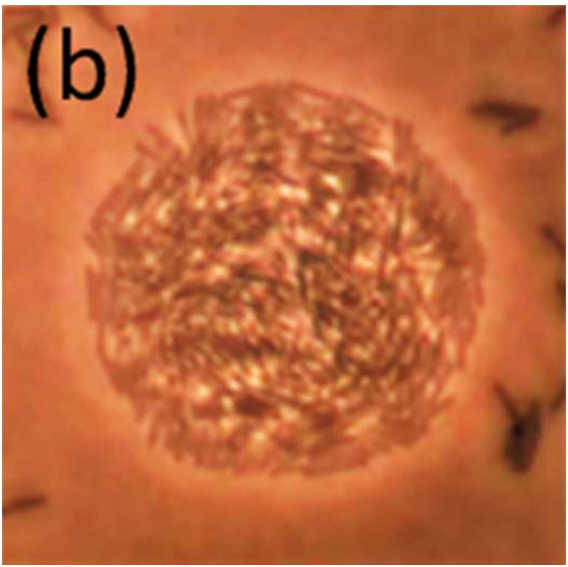

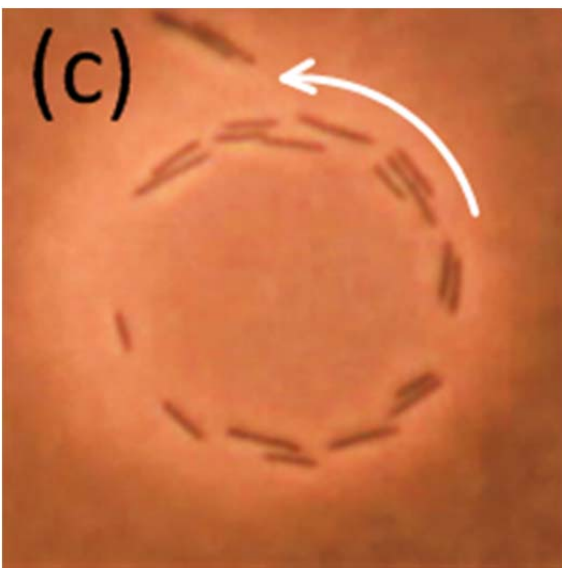
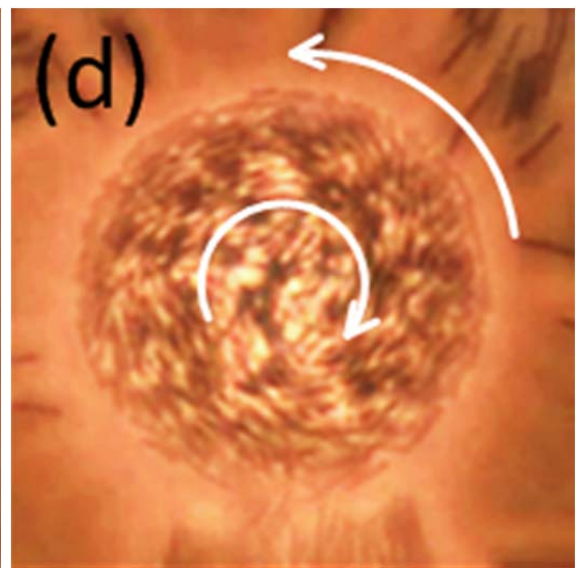

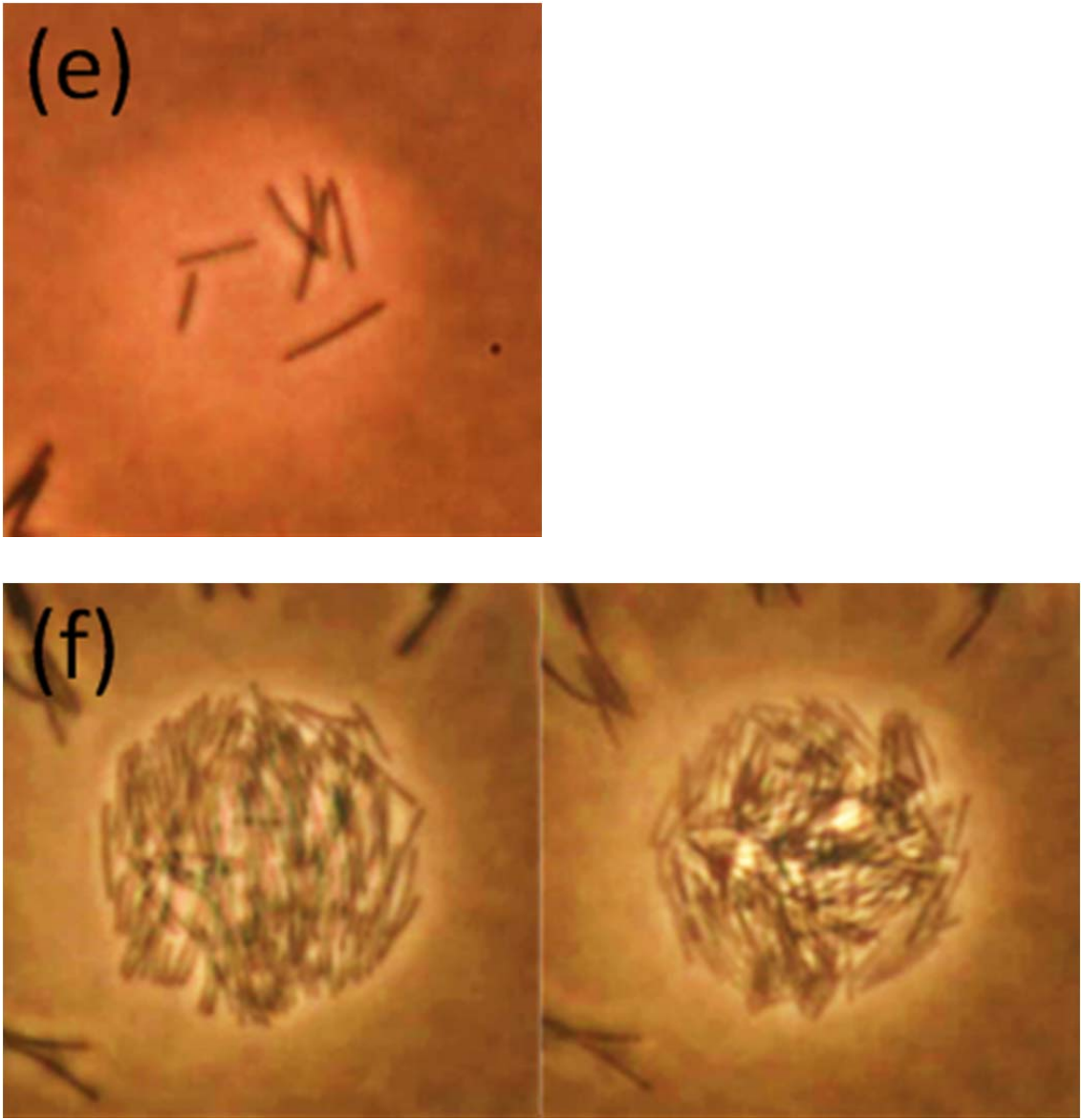

Fig. 2. (Color online) Snapshots of the six types of collective motion viewed from above the pool. The widths of the pictures are 60 μm. (a) Random motion: $l = 2.7$ μm, $\rho = 0.15$. (b) Turbulent motion: $l = 3.2$ μm, $\rho = 0.91$. (c) One-way rotational motion: $l = 5.8$ μm, $\rho = 0.07$. The arrow indicates the rotational direction of cell movements along the brim of the pool. (d) Two-way rotational motion: $l = 4.8$ μm, $\rho = 0.84$. The bacterial cells in the outer region swim counterclockwise along the brim of a pool, whereas those in the inner region swim clockwise as indicated by two arrows. (e) Random oscillatory motion: $l = 10.2$ μm, $\rho = 0.09$. (f) Ordered oscillatory motion: $l = 9.2$ μm, $\rho = 0.89$. Left and right pictures show ordered and disordered states, respectively. The interval time between these two states is about 10 min.

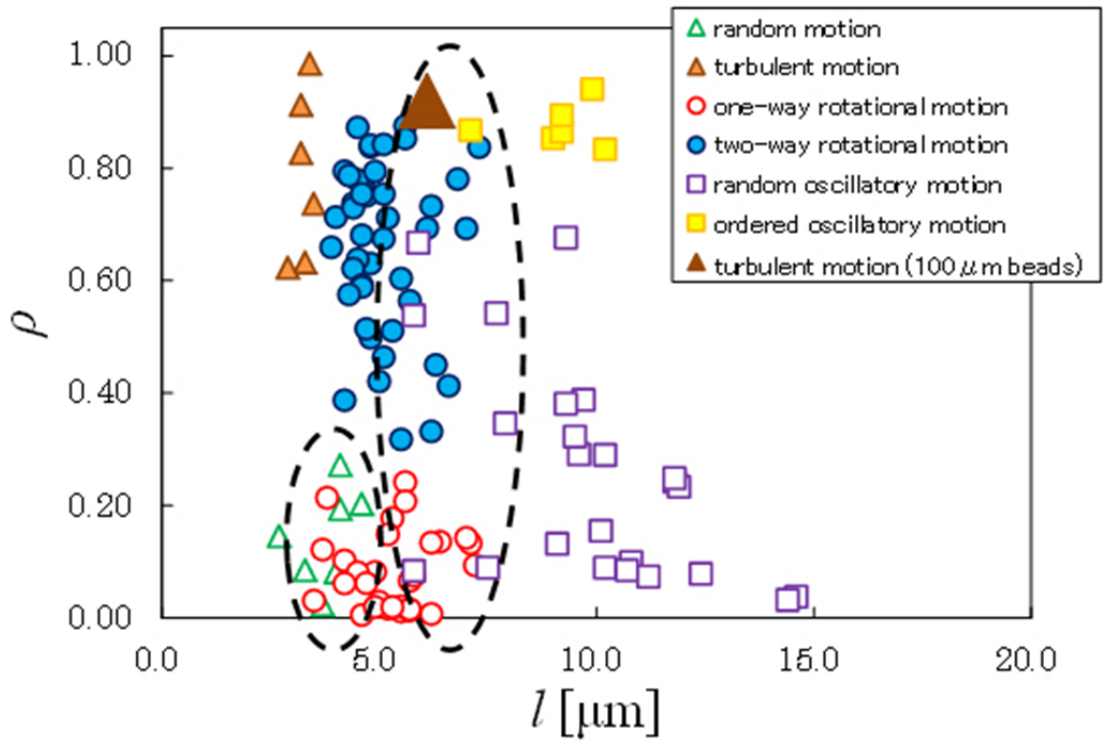

Fig. 3. (Color online) $(l, \rho)$ plots of collective motion of bacterial cells in a shallow circular pool. Open triangle: random motion. Filled triangle: turbulent motion. Open circle: one-way rotational motion. Filled circle: two-way rotational motion. Open square: random oscillatory motion. Filled square: ordered oscillatory motion. The large triangle indicates the turbulent motion observed in a large circular pool that is made by using a glass bead of 100 μm diameter. The different types of symbol overlap each other inside the dashed circles.

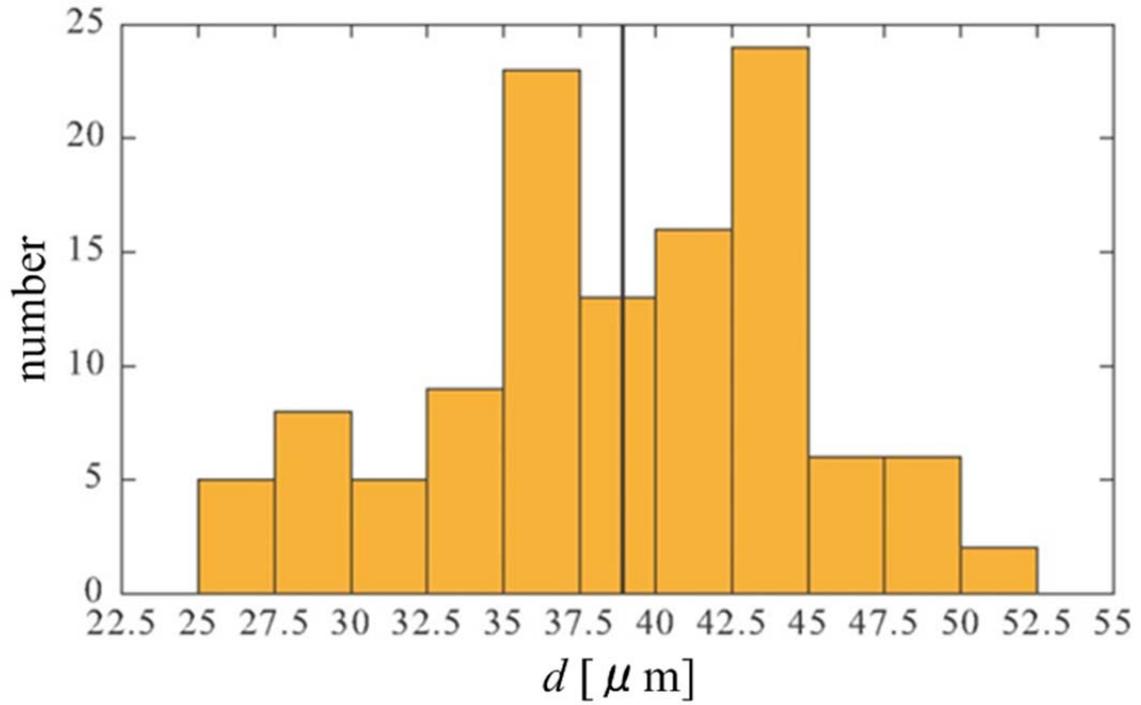

Fig. 4. (Color online) Distribution of the diameters $d$ of the 117 circular pools. The solid line indicates the average of $d$, which is 39 μm. The standard deviation is 6 μm.

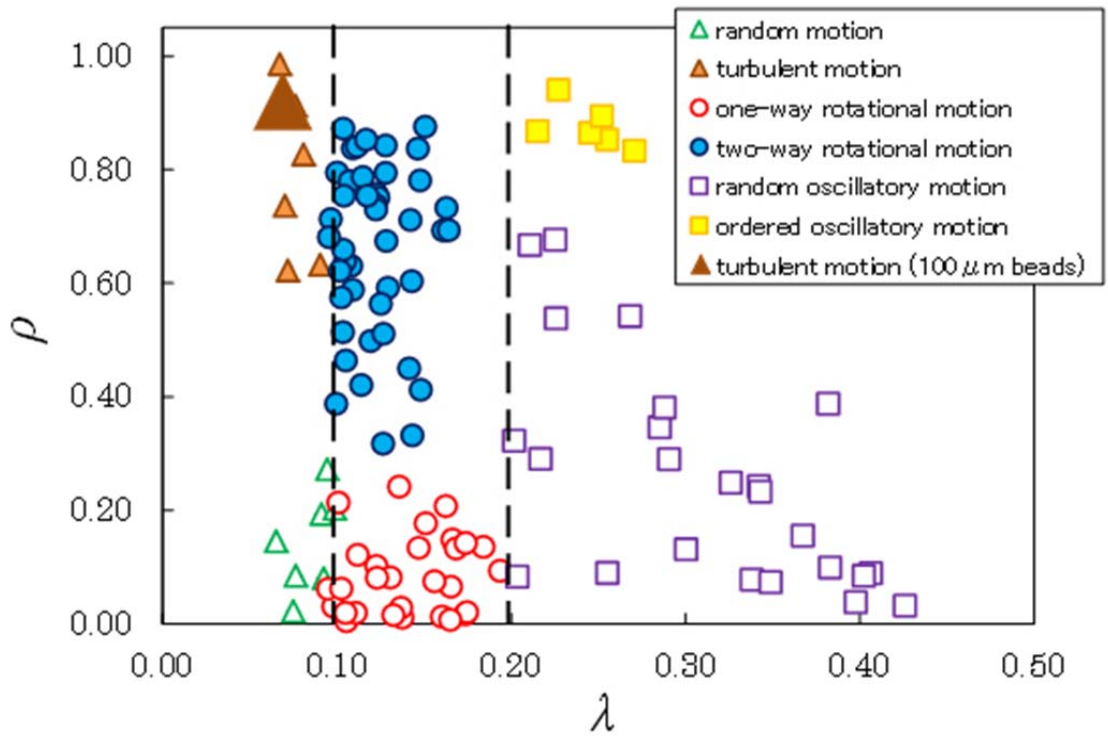

Fig. 5. (Color online) $(\lambda, \rho)$ plots of collective motion of bacterial cells in a shallow circular pool. The overlapping areas in Fig. 3 disappeared. The dashed lines indicate two critical values, $\lambda_{C1} = 0.1$ and $\lambda_{C2} = 0.2$. The symbols discriminating the six types of motion are the same as those in Fig. 3.

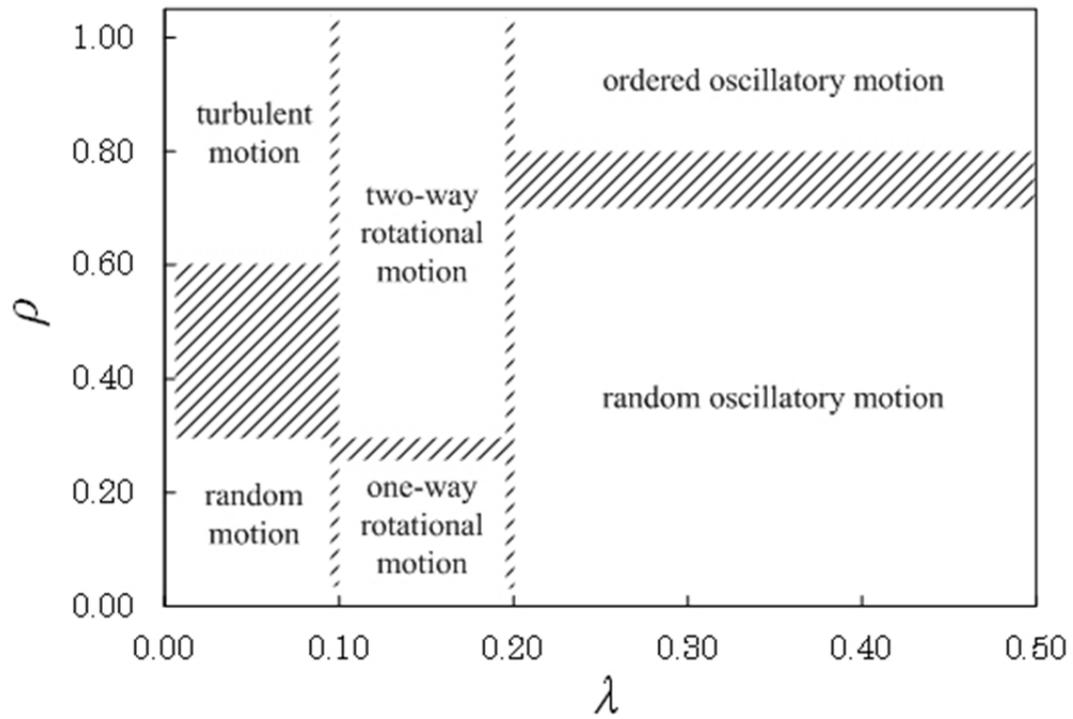

Fig. 6.  Phase diagram of the collective motion of bacterial cells in a shallow circular pool. The critical values of $\lambda$ are $\lambda_{C1} = 0.1$ and $\lambda_{C2} = 0.2$.